\documentclass[11pt, journal, draftcls, onecolumn]{IEEEtran}

\usepackage{enumerate}
\usepackage{amsmath, amsthm, amssymb}
\usepackage{float}
\usepackage{subcaption}
\usepackage{flexisym}
\usepackage{mathtools}
\usepackage{breqn}
\usepackage{array}
\usepackage{tikz}
\usepackage{cite}
\usetikzlibrary{shapes.geometric, arrows}
\usepackage[font = small]{caption}
\usepackage{paralist}
\usepackage{mathrsfs}
\usepackage{algorithm}
\usepackage{algpseudocode}
\usepackage[inline]{enumitem}
\usepackage{bbm}
\allowdisplaybreaks[1]

\DeclareMathOperator*{\argmin}{arg\,min}

\renewcommand{\IEEEQED}{\IEEEQEDclosed}

\begin{document}
\title{The Internet of Things: Secure Distributed Inference}
\author{Yuan Chen, Soummya Kar, and Jos\'{e} M. F. Moura
\thanks{Yuan Chen {\{(412)-268-7103\}}, Soummya Kar {\{(412)-268-8962\}}, and Jos\'{e} M.F. Moura {\{(412)-268-6341, fax: (412)-268-3890\}} are with the Department of Electrical and Computer Engineering, Carnegie Mellon University, Pittsburgh, PA 15217 {\tt\small \{yuanche1, soummyak, moura\}@andrew.cmu.edu}}
\thanks{This material is based upon work supported by the Department of Energy under Award Number DE-OE0000779 and by DARPA under agreement numbers DARPA FA8750-12-2-0291 and DARPA HR00111320007. The U.S. Government is authorized to reproduce and distribute reprints for Governmental purposes notwithstanding any copyright notation thereon. The views and conclusions contained herein are those of the authors and should not be interpreted as necessarily representing the official policies or endorsements, either expressed or implied, of DARPA or the U.S. Government.}}
\maketitle

\begin{abstract}
	The growth in the number of devices connected to the Internet of Things (IoT) poses major challenges in security. The integrity and trustworthiness of data and data analytics are increasingly important concerns in IoT applications. These are compounded by the highly distributed nature of IoT devices, making it infeasible to prevent attacks and intrusions on all data sources. Adversaries may hijack devices and compromise their data. As a result, reactive countermeasures, such as intrusion detection and resilient analytics, become vital components of security. This paper overviews algorithms for secure distributed inference {\color{black} in IoT}. %We focus on reactive countermeasures for consensus+innovations distributed estimators. Within the consensus+innovations framework, we present algorithms for adversary detection and resilient parameter estimation. 
\end{abstract}

\section{Introduction}
As the number of devices connected to the Internet of Things (IoT) continues to grow, the security of data generated, processed, and transceived by these devices becomes a pressing issue. IoT applications feature connected heterogeneous devices that share a common overarching goal; they cooperate and exchange information to complete their objective. For example, in a Smart Home, a car may communicate with the garage to automatically open the door, and wearable gadgets may exchange information with smart thermostats and lighting fixtures to create a comfortable environment~\cite{IoTDevices}. Connected automobiles in vehicular networks use on-board sensor measurements to monitor road conditions, find open parking spaces, and estimate traffic patterns~\cite{ConnectedVehicles, RoadSurface}. In mobile crowdsensing, individuals use their smartphones to monitor noise levels in neighborhoods and estimate wait times for public transportation~\cite{Crowdsensing}. In the Smart Grid, smart electricity meters make real-time measurements of power demand and consumption and are vital components of optimal power dispatch~\cite{SmartGrid}. Smart cities are instrumented with sensors to observe traffic, monitor weather, and measure air quality~\cite{IoTDevices, AoT, AirQuality}.  %Environmental sensors monitor air quality for concentrations of pollutants in cities and allergens and relay the information to inhabitants via smartphones and wearable gadgets~\cite{AirQuality}. %Each of these applicaitons relies on data-integrity: data that is untrustworthy leads to incorrect inferences.

{\color{black} Certain IoT applications feature devices specifically for monitoring and controlling physical systems. Machines on an assembly line may be equipped with sensors to detect production anomalies and predict when parts need to be replaced~\cite{IoTDevices}. Phasor measurement units, smart electricity meters, circuit breakers, and generation sources monitor and control the state of the Smart Grid~\cite{SmartGrid, SmartGridSurvey}. Other applications simply feature a collection of devices that need to cooperate to complete a shared task. For example, in a Smart Home, a smart speaker (e.g., Google Home, Amazon Echo) is responsible for deciphering a user's voice commands and relaying this information to a television, for broadcasting a movie, to lighting fixtures, for dimming the lights, or to the thermostat, for changing the temperature. The common characteristic of all of these IoT applications is that the devices must cooperatively process information to accomplish their collective objective, whether it is monitoring a physical system or dimming light features using voice commands. } 
\begin{figure}[h!]
\centering
\includegraphics[width = .75\columnwidth]{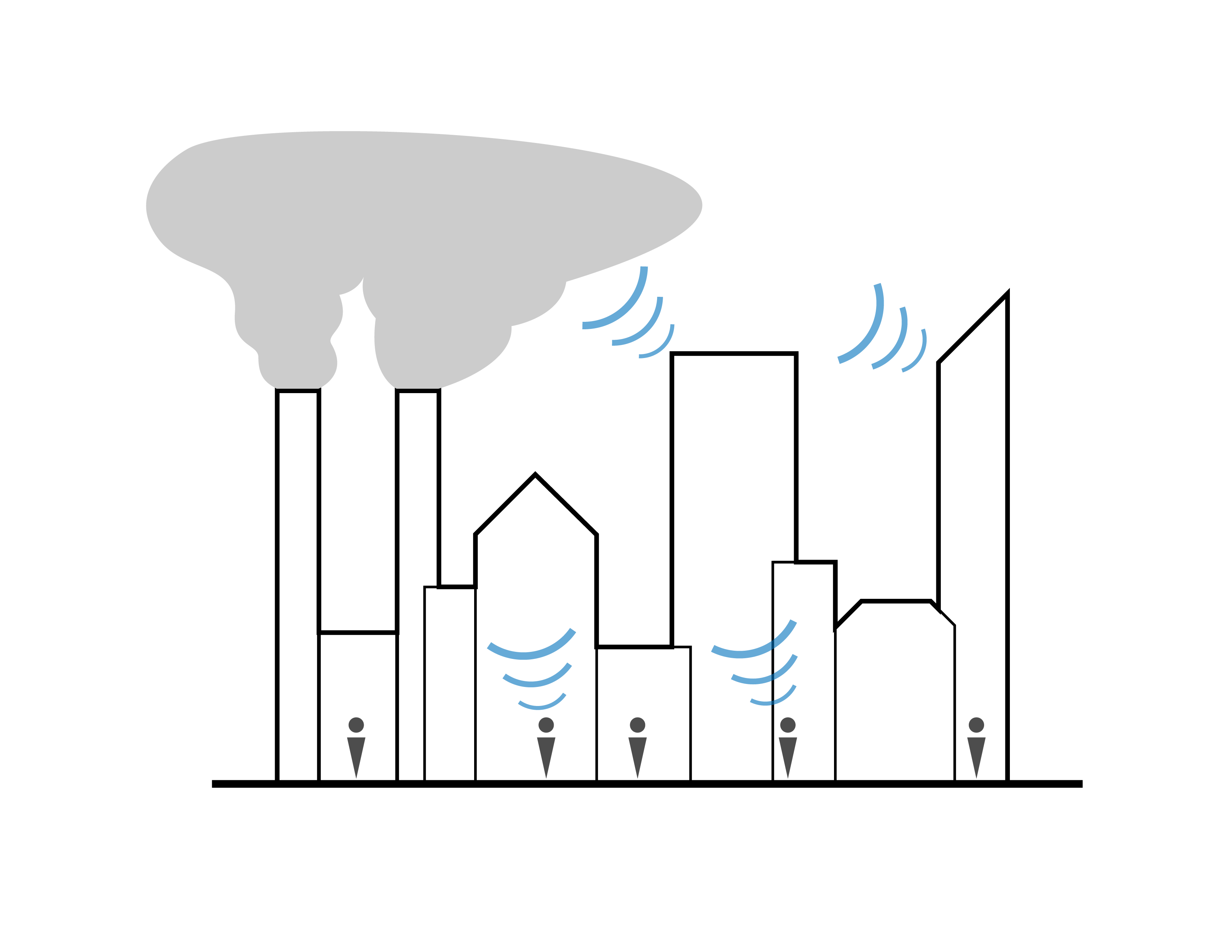}
\caption{An example IoT application is air quality monitoring. In air quality monitoring, a network of distributed sensors make measurements of local pollutant concentrations across the city. This information is then processed to produce a heat map of air quality over the city. Users may access this information via smartphones and wearable gadgets, see, e.g.,~\cite{AirQuality}.}
\end{figure}

A key task in these applications is inference, processing measurement data for information. The quality of inference critically depends on the integrity of the sensor data, i.e., on the trustworthiness of the sensors and devices producing the data. IoT devices, ranging in scale from pacemakers, to cars, to PMUs in the Smart Grid are vulnerable to cyber attack~\cite{IoTDevices, HACMS, SmartGridSurvey}. Malicious adversaries may hijack devices, arbitrarily corrupt their data, jam communication links, and mislead the application to produce erroneous inferences.

In this paper, we overview secure inference for IoT and highlight cooperation strategies that are resilient to data integrity attacks. Previous work has surveyed data analysis in IoT without adversaries~\cite{IoTData}, reviewed protocols for secure data communication~\cite{SecuritySurvey}, and summarized security challenges in IoT~\cite{DistributedSecurityChallenges}. Reference~\cite{SmartGridSurvey} surveys recent advances in security for the Smart Grid and presents a broad summary of secure data acquisition, communication, storage, and processing. In contrast, this paper provides a focused, more detailed overview of secure data processing and inference for IoT.

We consider three main architectures of IoT systems. In centralized architectures, a single entity processes all of the data from all of the devices. In decentralized or parallel architectures, devices perform local processing and transmit the processed data to a fusion center, which completes the computation task~\cite{DecentralDetection}. In fully distributed architectures, individual devices cooperate with neighbors over a communication network and perform all of the processing. We present algorithms for each of these architectures that fuse data streams from many separate devices and still produce a collective accurate inference even while an adversary tampers with a subset of the devices and manipulates their data streams. 

The end goal is to correctly process the data even in the presence of adversaries. One strategy to achieve this goal is by detecting and identifying attacks and, after doing so, taking corrective action. An alternative approach is to use resilient processing algorithms, which, by design, resist attacks without explicit detection and identification.  This paper overviews both types of strategies. The techniques we present are generic and \textit{not} application specific; for illustrative purposes, however, we will explain these techniques in the context of air quality monitoring. In practice, one is interested in graceful degradation of performance in the presence of adversaries. The results and approaches we consider here, due to lack of space, focus on strategies with performance assurances that either guarantee success or signal compromised assets.

\subsection{Preventive and Reactive Security}

To protect data integrity, security countermeasures fall into two main categories: preventive countermeasures and reactive countermeasures. Preventive countermeasures seek to \textit{prevent} intrusion attempts by directly protecting data and communications~\cite{SecuritySurvey}. Examples of preventive security include cryptographic protocols to authenticate the identity of devices and to authorize users to access data. Authentication and authorization protocols ensure that a fusion center receives data streams only from trusted devices. They prevent an adversary from introducing malicious data to the fusion center via a rogue, unverified device~\cite{Authentication}. 

Reactive countermeasures aim to mitigate failures in preventive security and ensure that the system continues to operate properly even when preventive security breaks down~\cite{SecuritySurvey}. Whereas preventive security protects IoT systems by making it more difficult for an adversary to compromise data and devices, reactive security ensures that systems operate resiliently even when a number of devices become hijacked. Reactive countermeasures include attack detection~\cite{Liu, ChenTAC} and identification~\cite{Fawzi} algorithms for cyber-physical systems (CPS). The objectives of attack detection and identification are, respectively, to determine if the measurements from any of the sensors have been altered by an adversary and to identify specifically which sensors have been compromised. After detecting or identifying an attack, the system may take corrective action to mitigate the damage~\cite{Fawzi}. 

\textit{Remark}: {\color{black} CPS, cyber-physical systems, refer to \textit{physical} systems instrumented with a layer of cyber devices. Examples of CPS include robotic platforms, drones, and modern automobiles. These are physical systems that are highly instrumented with sensors and actuators (e.g., devices to measure and control speed and acceleration)~\cite{HACMS}. At the other extreme of scale in CPS, large infrastructures, like the power grid, are also highly instrumented by sensors and actuators (e.g., phasor measurement units, smart meters, circuit breakers, and generation sources in the power grid). These devices provide the CPS, be it the modern automobile or the power grid, with a cyber-layer. The devices are not necessarily themselves interconnected, but their measurements are processed by a central processing unit (CPU), like in an automobile, or by a supervisory control and data acquisition (SCADA) center, like in the power grid. The important characteristic of CPS is that there is an underlying physical system (e.g., an automobile or the power grid) instrumented by a cyber-layer.

An IoT is more generally a panoply of devices instrumenting, for example, the refrigerator, oven, lighting fixtures, and other appliances in a smart kitchen, or, connecting wearables, smartphones, and smart speakers to a television~\cite{IoTDevices}. In the IoT, the network provides the ability for the heterogeneous devices to cooperate to achieve a common task. For example, a smart speaker deciphers voice commands and relays them to television to show a specific video, and, of course, in the Smart Grid, the IoT devices monitor the state of the grid. }
%There IoT devices that are not considered CPS. For example, connected smartphones and wearable gadgets are IoT devices, and they may gather and share information with each other~\cite{WhatIsIoT}, but these devices are not cyber-physical systems. %IoT is a network of sensors or devices. CPS are physical systems such as driver-less cars, drones, mobile robotic platforms, and the Smart Grid, and they may be instrumented with devices, e.g., phasor measurement units (PMUs) in the Smart Grid. If these devices are networked, then they provide an example of an IoT. %CPS may be IoT devices themselves (e.g., drones and mobile robotic platforms) or may be comprised of IoT devices (e.g., smart meters in the electricity grid). There are IoT devices that are not considered CPS, e.g., smartphones, wearable devices; the key feature that distinguishes CPS from other IoT devices is the underlying physics of the CPS. 
\hfill $\small\blacksquare$

We further classify reactive countermeasures as either explicit or implicit. Explicit countermeasures directly detect and identify malicious behavior, to alert the system to compensate for adversarial activity. For example, reference~\cite{Fawzi} designs an attack identification algorithm for cyber-physical systems (CPS). In a CPS, such as a remotely controlled vehicle, we are interested in estimating the state of the system (e.g., position, velocity, etc.) from its onboard sensor measurements (e.g., odometer, accelerometer, GPS, camera, lidar). An adversary may mislead the state estimator by altering some of the sensor measurements~\cite{HACMS}. Reference~\cite{Fawzi} provides a method to identify the compromised sensors and determine the amount by which the measurements were altered. Attack detection and identification, by themselves, may not mitigate the attack. Still, they are important components of secure data processing; once an attack is correctly detected and identified, the IoT system can take corrective actions. For example, in state estimation~\cite{Fawzi}, once the attack has been identified, the state estimator can compensate for the altered measurements and recover the system state.

Implicit countermeasures do not alert IoT systems to malicious behavior. Instead, they provide resilience by limiting the impact of malicious behavior on the system's end goal. As an example, consider a network of sensors monitoring air quality. Individual sensors maintain estimates of global pollutant concentration levels using their measurements of local pollutant concentrations and exchanging estimates with neighboring sensors. Sensors update their estimates as a weighted average of their own previous estimates, their neighbors' previous estimates, and their local measurements. To limit the impact of adversely corrupted measurements, a sensor assigns lower weight to local measurements that deviate more from its estimate~\cite{ChenDistributed2}. This method may not identify which sensors have been attacked but nevertheless provides resilience against adversarial devices and ensures that the remaining sensors successfully estimate global pollutant concentrations.

%As an example, consider again a fully distributed architecture for air quality monitoring. Individual sensors maintain estimates of global pollutant concentration levels using their measurements of local pollutant concentrations and exchanging estimates with neighboring sensors. Sensors update their estimates as a weighted average of their own previous estimates, their neighbors' previous estimates, and their local measurements. To limit the impact of adversely corrupted measurements, a sensor assigns lower weight to local measurements that deviate more from its estimate~\cite{ChenDistributed2}. This method may not identify which sensors have been attacked but nevertheless provides resilience against adversarial devices and ensures that the remaining sensors successfully estimate global pollutant concentrations.

\subsection{IoT Architectures and Security Countermeasures}
Countermeasure techniques against data manipulating adversaries depend on the architecture of the IoT application. In particular, they depend on whether or not the architecture contains a central processor. In centralized architectures, a single processor, possibly in the cloud, has access to data from all devices. For example, in a centralized air quality monitoring system, the central estimator has access to local measurements of pollutant concentrations from all sensors placed throughout a city and fuses the data to produce a heat map of air quality over the entire city. In this architecture, the individual devices perform minimal processing, and there is no device-to-device communication. %In traditional decentralized architectures, individual end devices (e.g., wearable gadgets, smartphones, electricity meters, air pollutant sensors) collect measurements and data. The devices then transmit, say, through access points or base stations and the backbone telecom network, their data or local decisions to a data fusion center that completes the inference task, usually in the cloud~\cite{IotData, DistributedSecurityChallenges, ConnectedVehicles2}. 
%The fusion center then relays the result back to the end devices and users. For example, in instrumenting a decentralized air quality monitoring system, sensors placed throughout a city each make local measurements of pollutant concentrations and transmit them to a fusion center. The fusion center then fuses the data and produces a heat map of air quality over the entire city. Finally, it transmits the air quality information back to a user's end device. . The fusion center coordinates the communication with all of the devices and performs the computations. 

%The combination of preventive and reactive countermeasures is suitable for the traditional, cloud-based, centralized architecture for IoT. In the cloud IoT architecture, individual smart devices communicate their data to a data fusion center, and all of the computations necessary for inference are performed at the fusion center~\cite{IotData, DistributedSecurityChallenges, ConnectedVehicles2}. For security purposes, the fusion center authenticates the identities of devices transmitting the data, encrypts the data to prevent unauthorized access, and employs reactive (centralized) inference algorithms that are resilient to breaches in preventive security. The authentication and encryption algorithms require high computational power and the centralized resilient inference algorithms require access to data from all connected devices, both of which are available at a data fusion center.

The growing \textit{intelligence} of IoT devices enables edge, fog, or micro-edge computing, where some of the computational burden is offloaded from the central processor to the end devices~\cite{Fog}. Like a centralized architecture, a decentralized architecture also features a central processor (i.e., a fusion center) that collects information from individual devices. In decentralized architectures, there is no device-to-device communication. The difference between centralized and decentralized architectures is that, in a decentralized edge architecture, individual devices may process the local data or produce local decisions that are then transmitted to the fusion center~\cite{DecentralDetection}. For example, a sensor may transmit a time moving averaged, quantized version of its data to the fusion center~\cite{FusionCenterEstimation1}. Decentralized architectures may also be referred to as parallel architectures.

In addition to fog or micro-edge computing, IoT applications {\color{black} will increasingly} involve device-to-device communication: that is, instead of transmitting and receiving data to and from the cloud or edge, end devices may communicate directly with each other. The combination of device computing and device-to-device communications enables fully distributed web-like IoT architectures. {\color{black} We will assume in this paper that} in a fully distributed architecture, there is no fusion center and end devices carry all of the computational burden. Devices make local measurements, exchange information with neighboring devices, and process their available information to perform inference. In distributed air quality monitoring, each sensor will update its estimate pollutant concentrations and will communicate it to neighboring sensors. Through this cooperation, users' end devices obtain air quality about the city, as we will explain when we discuss distributed estimation.

Edge, fog, or micro-edge computing architectures are better suited than cloud architectures for applications with low latency and real-time processing requirements~\cite{Fog}. Furthermore, the distributed architecture is well suited when there is no central coordinator. One example is vehicular networks in which individual vehicles may exchange information with other nearby vehicles to determine traffic information for path planning or to coordinate merging and lane changes~\cite{ConnectedVehicles}. To accommodate the dynamic environment, vehicular driving is computed in real-time, so, {\color{black} latency looms as a large issue, and} computations should be performed at the vehicle rather than in the cloud.

The combination of preventive and reactive countermeasures is suitable for cloud-based IoT architectures with central processors. The authentication and encryption algorithms require high computational power and the centralized resilient inference algorithms require access to data from all connected devices, both of which are available at a data fusion center. Compared to architectures with central processors, security (in particular, data integrity) in distributed architectures faces more difficult challenges~\cite{DistributedSecurityChallenges}. Individual devices lack the computational capabilities of a cloud data center and may not be able to implement all of the same preventive security measures. In architectures with central processors, preventive countermeasures aim to secure a single entity. In comparison, a distributed architecture consists of numerous devices that are deployed in many different physical locations. Thus, it may be infeasible for preventive security measures to protect all of the devices. Moreover, in a distributed architecture, individual devices do not have access to data from \textit{all} other devices, and, as a result, they are not able to execute the centralized resilient inference algorithms used in fusion centers.

\begin{figure}[h!]
\centering
\begin{subfigure}[h]{0.48\columnwidth}
	\includegraphics[width = \columnwidth]{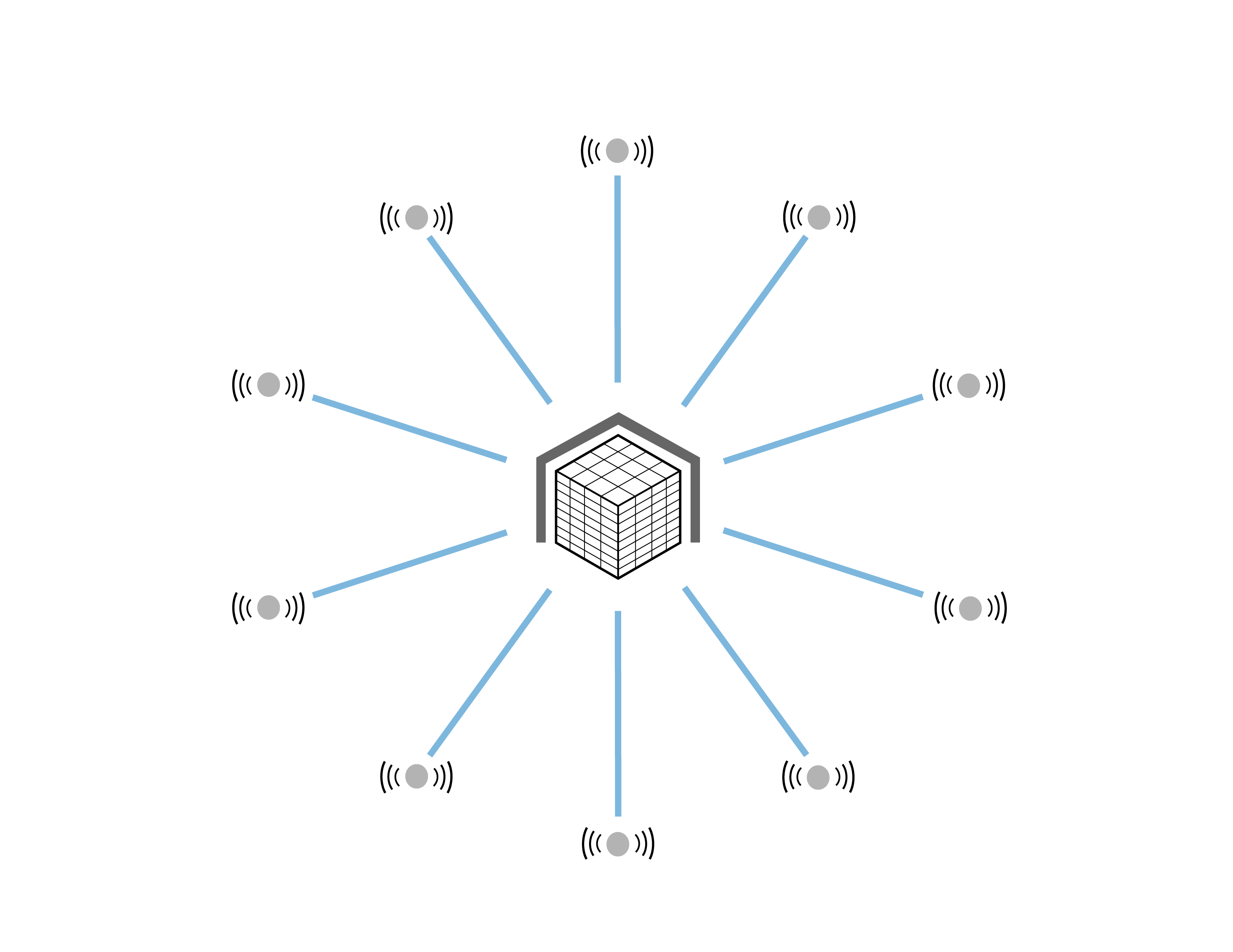}
	\caption{Architecture with Central Processor}\label{fig: decent}
\end{subfigure}
\begin{subfigure}[h]{0.48\columnwidth}
	\includegraphics[width = \columnwidth]{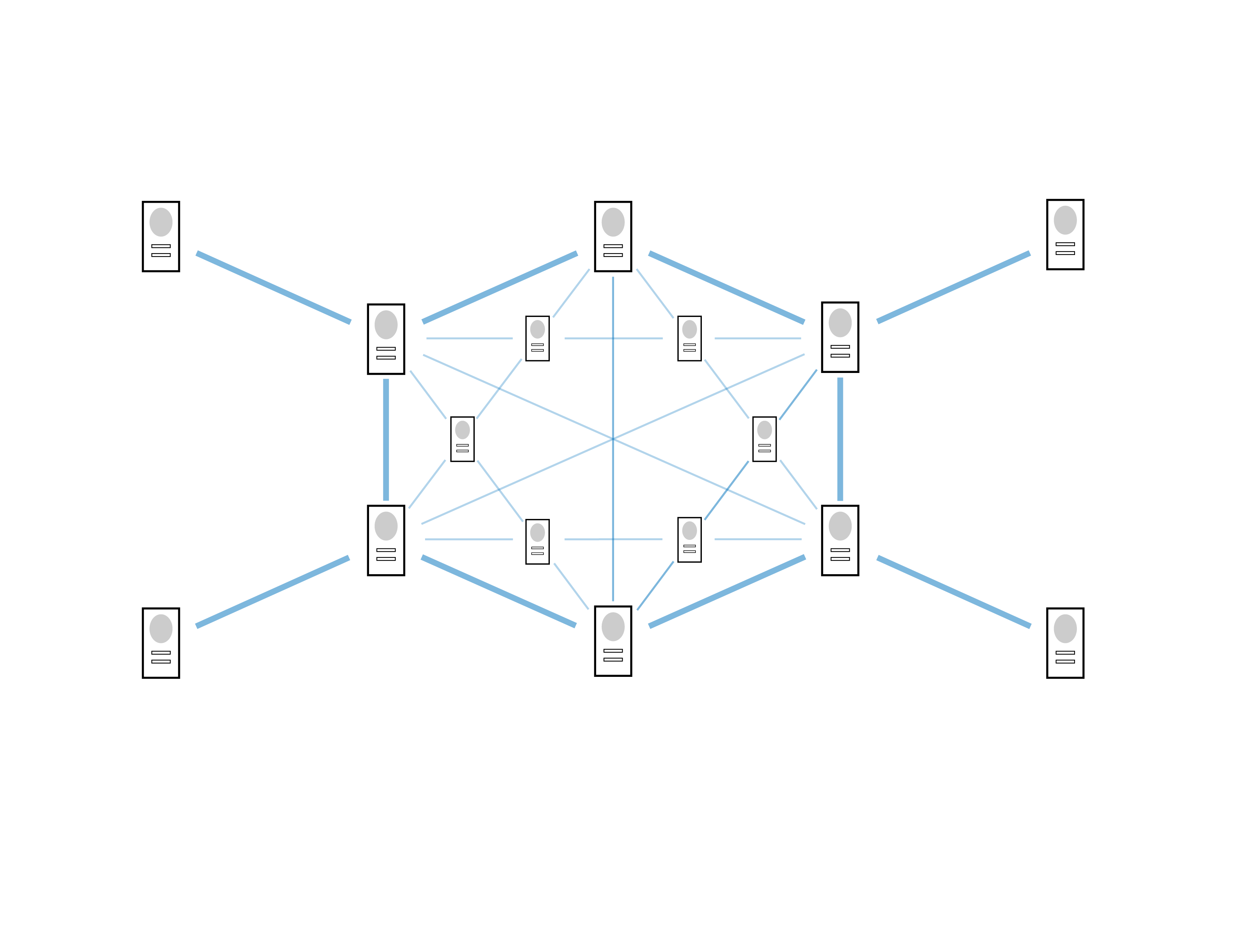}
	\caption{Fully Distributed Architecture}\label{fig: dist}
\end{subfigure}
\caption{(a) Devices transmit raw or processed data to a central processor in centralized and decentralized (or parallel) architectures (see, e.g.,~\cite{DecentralDetection, FusionCenterEstimation1, Liu, ChenTAC, Fawzi, Tsitsiklis}). There is no device-to-device communication. (b) In distributed architectures, devices communicate with each other and cooperate to complete inference tasks without a fusion center (see, e.g.,~\cite{DistributedSecurityChallenges, TBA, Boyd1, QuantConsensus, ChenDistributed1, ChenDistributed2}).}\label{fig: architectures}
\end{figure}

\section{General Measurement and Attack Model}
We model an IoT system as a collection of agents or devices $\left\{1, 2, \dots, N\right\}$, each measuring an unknown parameter $\theta^*$. Individual devices make local, noisy measurements of the phenomenon of interest~\cite{Liu, FusionCenterDetection2, Varshney2, FusionCenterEstimation1, Varshney1, ChenDistributed1, DiffusionLMS2}. The measurement $y_n(t)$, of the $n^{\text{th}}$ agent is
\begin{equation}\label{eqn: measurementModel}
	y_n(t) = f_n(\theta^*) + n_t,
\end{equation}
where $t$ is time, $f_n$ is the local measurement function, and $n_t$ is the measurement noise. The function $f_n$ is nonlinear in general. An example of nonlinearity is sensor saturation; physical sensors have maximum and minimum measurement levels and exhibit saturation and clipping when the parameter of interest, $\theta^*$, exceeds these bounds. The agents' goal is to recover the parameter $\theta^*$ from their measurement streams.

%\subsection{Byzantine Adversaries}
An adversary may hijack a subset of the agents and manipulate their data streams. A standard classical motivation for data integrity attacks against IoT systems comes from the Byzantine Generals Problem, where a group of generals decides whether or not to attack a city by passing messages among one another (in an all-to-all manner)~\cite{ByzantineGenerals}. Traitor generals attempt to mislead the remaining loyal generals by sending false messages. The authors of~\cite{ByzantineGenerals} provide an algorithm for the loyal generals to reach the correct decision (using an all-to-all communication setup) even when up to one third of the generals are traitors. The classical paradigm of all-to-all communication has been relaxed in, for example, decentralized setups where a fusion center combines all received local decisions (including those of the traitors). In IoT applications, a Byzantine adversary (or Byzantine device) is a device that attempts to disrupt an inference task by transmitting falsified data.

\section{Secure Inference with a Central Processor}
\subsection{State Estimation in Cyber-Physical Systems}
The IoT can monitor critical infrastructure and cyber-physical systems (CPS), such as the electricity grid and autonomous vehicles. Electricity meters measure power consumption levels and operating conditions of the Smart Grid~\cite{SmartGrid}. A vital task in operating the Smart Grid is state estimation, i.e., determining the voltage angles and magnitudes at each bus in the grid from meter measurements. Reference~\cite{Liu} studies state estimation for power grids when an adversary manipulates a subset of the measurements. In~\cite{Liu}, a  fusion center collects measurements from all of the meters, following the linearized measurement model
\begin{equation}\label{eqn: gridMeasurement}
	\mathbf{y} = \mathbf{H}\theta^* + \mathbf{w} + \mathbf{a}.
\end{equation}
In~\eqref{eqn: gridMeasurement}, $\mathbf{y}$ is the collection of measurements from the meters (each component of $\mathbf{y}$ is a measurement from a single meter), $\theta^*$ is the state of the grid (i.e., the voltage angles and magnitudes at each of the buses), $\mathbf{H}$ is a matrix that describes what each meter measures, $\mathbf{w}$ is the measurement noise, and $\mathbf{a}$ models the adversarial attack. {\color{black} In this section, we consider a centralized architecture where a central processor has access to measurements from all of the sensors.}

The adversary manipulates the data from a subset of the meters, and the components of $\mathbf{a}$ describe the amount by which a particular measurement is changed. If a meter is not under attack, then, the corresponding component of $\mathbf{a}$ is $0$. If only a few sensors are attacked, $\mathbf{a}$ is a sparse vector whose nonzero entires may have arbitrary values as determined by the adversary. The goal in state estimation is to recover the value of $\theta^*$ from the measurement $\mathbf{y}$. To deal with the adversary,~\cite{Liu} proposes an attack detector. First, the system solves the optimization problem
\begin{equation}\label{eqn: gridEstimate}
	\widehat{\theta} = \argmin_\theta \left\lVert \mathbf{y} - \mathbf{H} \theta \right\rVert_2,
\end{equation}
that is, the system finds the estimate $\widehat{\theta}$ that minimizes the squared error between the measurement $\mathbf{y}$ and the predicted measurement $\mathbf{H} \widehat{\theta}$. Then, the attack detector declares that an attack has occurred if the energy of the estimation residual $\mathbf{y} - \mathbf{H} \widehat{\theta}$ exceeds a threshold $\tau$, i.e., if $\left\lVert \mathbf{y} - \mathbf{H} \widehat{\theta} \right\rVert_2 > \tau$. This algorithm fails to detect certain attacks. As the authors of~\cite{Liu} show, the adversary can avoid being detected by compromising meters in a specific way such that $\mathbf{a}$ belongs to the column space of the measurement matrix $\mathbf{H}$.

\textit{Remark}: Resilience in parameter estimation tasks depends on observability. A model is observable if, in the absence of noise and attacks, it is possible to exactly recover the parameter of interest from all the sensor measurements; otherwise, it is unobservable. For the model~\eqref{eqn: gridMeasurement} {\color{black} and centralized architectures}, observability means that the matrix $\mathbf{H}$ has full column rank, so there is a unique value of the parameter $\theta^*$ that corresponds to any value of the noiseless measurement $\mathbf{H} \theta^*$. {\color{black} We will discuss the notion of observability for distributed architectures in the sequel.}The algorithm in~\cite{Liu} detects attacks on up to $s$ sensors as long as the measurement model~\eqref{eqn: gridMeasurement} is observable after removing any $s$ sensors. \hfill $\small\blacksquare$

Reference~\cite{Liu} addresses detecting measurement attacks against \textit{static} state estimation, i.e., estimating a parameter that does not change over time. For CPS such as autonomous vehicles, we are interested in estimating a \textit{dynamic} parameter that changes over time. For example, in context of the measurement model~\eqref{eqn: gridMeasurement}, for an autonomous vehicle, the state $\theta_t$ describes its position and velocity at time $t$, and $\theta_t$ evolves over time depending on the vehicle's physics. Onboard sensors, such as odometers and GPS, measure the state $\theta_t$ and communicate the data to the vehicles data fusion center. The goal of the fusion center is to recover $\theta_t$, the vehicles current position and velocity, from the sensor measurements. Just like in the power grid, an adversary can alter the measurement data from some of the vehicle's sensors.

Reference~\cite{ChenTAC} proposes for dynamic CPS, like autonomous robots, a detector for attacks on sensor measurements.  The dynamic attack detector in~\cite{ChenTAC} is similar to the static detector in~\cite{Liu}. The key difference is that~\cite{ChenTAC} uses a model that accounts for a system's dynamics (e.g., laws of physics that describe the motion of a vehicle and bounds on its acceleration) and maps a system state to a \textit{sequence} of predicted measurements over time. Following this model, the detector from~\cite{ChenTAC} collects a sequence of sensor measurements over time, computes a state estimate, and reports an attack if the energy of the estimate residual (the difference between the observed sensor measurements and the sensor measurements predicted from the state estimate) exceeds a certain threshold. This algorithm detects all sensor attacks so long as the system with only the uncompromised sensors is observable.

The authors of~\cite{Fawzi} go beyond attack detection and provide an algorithm to identify the sensors under attack. The attack identification algorithm is applicable to both static and dynamic settings. In the context of power grid state estimation~\eqref{eqn: gridMeasurement}, the goal of attack identification is to recover the value of the attack $\mathbf{a}$ using the meter measurements $\mathbf{y}$ and the measurement matrix $\mathbf{H}$. The authors of~\cite{Fawzi} formulate the attack identification problem (in a static setting) as solving the optimization problem 
\begin{equation}\label{eqn: attackIdentificationOpt}
	\widetilde{\theta} = \argmin_{\theta} \left\lVert \mathbf{y} - \mathbf{H} \theta \right\rVert_0
\end{equation}
and recovering the attack $\mathbf{a}$ as $\widetilde{\mathbf{a}} = \mathbf{y} - \mathbf{H}\widetilde{\theta}$.\footnote{Reference~\cite{Fawzi} assumes a noiseless measurement model, which means, in the context of~\eqref{eqn: gridMeasurement}, that $\mathbf{w} = 0$.} The idea behind~\eqref{eqn: attackIdentificationOpt} is that the adversary can only change the measurements on a few sensors, so the corresponding attack vector $\mathbf{a}$ contains mostly zeros with sparse nonzero elements. The attack identification algorithm finds the state $\widetilde{\theta}$ and the most sparse attack $\widetilde{\mathbf{a}}$ that explains the measurement $\mathbf{y}$. 

In~\eqref{eqn: attackIdentificationOpt}, the goal is to find the estimate $\widetilde{\theta}$ that is consistent with the highest \text{number} of sensor measurements. The \textit{amount} by which the observed measurement and predicted measurement (from $\widetilde{\theta}$) differ does not matter, since it is assumed that this difference comes as a result of adversarial attack. In contrast, in~\eqref{eqn: gridEstimate}, the goal is to find the estimate that minimizes the total squared error between the observed measurement and the predicted measurement. The algorithm in~\cite{Fawzi} identifies any attack on up to $s$ sensors if the measurement model~\eqref{eqn: gridMeasurement} is observable after removing any $2s$ sensors. The optimization problem in~\eqref{eqn: attackIdentificationOpt} is non-convex, and, to solve~\eqref{eqn: attackIdentificationOpt}, we must check all possible sets of attacked sensors~\cite{Fawzi}. The number of possible sets of attacked sensors increases exponentially as the total number of sensors increases. To make the problem tractable,~\cite{Fawzi} relaxes~\eqref{eqn: attackIdentificationOpt} by replacing the $\ell_0$ pseudo-norm with the $\ell_1$ norm. 

The attack detection~\cite{Liu, ChenTAC} and identification~\cite{Fawzi} algorithms for CPS are explicit countermeasures against attackers. Their objective is to alert the system to attacks against sensors so that it can take corrective actions to mitigate the effects of the attacks. It is easier to explicitly detect intrusions than it is to design resilient estimation algorithms. The drawback is that attack detection algorithms are incomplete solutions for secure inference, since, once an attack has been detected, the system still needs to take corrective action to mitigate the effects of the attack.

\subsection{Decentralized Hypothesis Testing}
In decentralized hypothesis testing, a group of $N$ sensors measures a phenomenon that falls under one of two hypotheses: $\mathcal{H}_0$ and $\mathcal{H}_1$, occurring with prior probabilities $P_0$ and $P_1$, respectively. The sensors communicate with a fusion center whose goal is to determine which hypothesis is true. For example, the sensors can measure environmental conditions in a factory~\cite{IoTData}, and the goal of the fusion center is to determine whether the conditions are safe ($\mathcal{H}_0$) or hazardous ($\mathcal{H}_1$) for the factory's workers. Due to communication constraints, the sensors do not transmit their measurements directly to the fusion center. Instead, each sensor decides, based on its local measurements, which hypothesis is true and transmits the local decision ($\mathcal{H}_0$ or $\mathcal{H}_1$) to the fusion center. In the absence of attacks, the decentralized hypothesis testing problem has been extensively studied~\cite{Tsitsiklis, Varshney3, Aldosari}.

In the presence of Byzantine attacks, an adversary compromises a fraction $\alpha$ of the sensors, and the Byzantine sensors transmit arbitrary decisions in order to mislead the fusion center. The authors of~\cite{FusionCenterDetection2} determine the minimum fraction $\alpha$ of Byzantine agents to ensure that the fusion center cannot distinguish between the two hypotheses. For vector observations, the adversary must compromise at least half of the sensors to ensure that the two hypotheses are indistinguishable~\cite{FusionCenterDetection2}. When the fraction of Byzantine sensors is less than one half, reference~\cite{Varshney2} designs a fusion rule that is resilient to Byzantines: the fusion center declares hypothesis $\mathcal{H}_1$ if at least $K^*$ of the sensors declare $\mathcal{H}_1$ locally. The threshold $K^*$ depends on the prior probabilities of $\mathcal{H}_0$ and $\mathcal{H}_1$ and the desired level of resilience, i.e., the fraction of Byzantines that need to be tolerated. Reference~\cite{SpecSensing} analyzes the effect of Byzantine agents in decentralized hypothesis testing in the context of collaborative spectrum sensing.

\subsection{Security with a Central Processor: Summary and Other Work}
In architectures with central processors, devices transmit local raw data or local decisions to the central processor, and adversarial devices transmit falsified data to disrupt the inference task. Both explicit and implicit countermeasures require that the uncompromised devices have enough influence to overcome the effects of adversarial behavior. For example, in CPS, the collection of uncompromised sensors must be observable in order to detect sensor attacks~\cite{ChenTAC}, and, in~\cite{Varshney2}, a majority of devices need to remain uncompromised in order for the fusion center to resiliently perform hypothesis testing.  

%Further work in decentralized resilience includes algorithms for resilient parameter estimation with quantized data~\cite{FusionCenterEstimation1} and methods to identify Byzantine devices in hypothesis testing~\cite{Varshney1}. 

Additional work in secure inference with central processors includes~\cite{FusionCenterEstimation1}, which provides an algorithm for resilient decentralized parameter estimation with quantized data. The authors of~\cite{Varshney1} design methods to identify Byzantine devices in hypothesis testing. Reference~\cite{SmartGridAttack} surveys attacker strategies and detection methods for data integrity attacks against the Smart Grid. In addition,~\cite{SmartGridAttack} proposes a method to detect attacks in the shortest amount of time (i.e., quickest detection). Reference~\cite{Jamming2} studies state estimation under jamming attacks: in jamming attacks, instead of manipulating data streams, the attacker prevents the devices from communicating with the central processor. The authors of~\cite{Jamming2} analyze jamming attacks against state estimators in a game-theoretic framework and find Nash equilibrium strategies for both the attacker and the estimator. 

\section{Secure Inference in Distributed Architectures}
In a fully distributed architecture, there is no fusion center to collect data from all of the devices. {\color{black} This is a simplistic architecture, since in the real world we may expect a hybrid or hierarchical architecture where devices communicate among themselves, with edge (intermediate computing resources), and the cloud. At different levels, different strategies may be used, including a mix of the ones we described in Section III and the ones we consider here. In a distributed architecture,} devices communicate with each other to complete computation and inference tasks. 

For simplicity, we consider a flat network of $N$ devices (or agents), $\left\{1, 2, \dots, N \right\}$. We model the communication between devices with an undirected simple graph $G = (V, E)$. For background on graphs, see~\cite{Spectral}. The vertex set $V$ of $G$ is the set of $N$ devices, and the edge set $E$ describes the communication links among them. Two devices are connected by an edge if they can communicate with each other. A device can only communicate with its neighbors in the graph $G$. The set $\Omega_n$ is the neighborhood of device $n$, i.e., the set of all devices that share a communication channel with device $n$. For example, for distributed air quality monitoring in a city, individual sensors may only communicate with nearby sensors, instead of communicating with all other sensors in the city.  We now consider two distributed inference tasks. The first is distributed consensus, where sensors or devices cooperatively compute a statistic of a snapshot of data, for example, the average of the (distributed) data. The second is distributed estimation where sensors cooperate iteratively process a stream of measurements to recover the value of an unknown parameter.  

\subsection{Resilient Consensus}
In consensus, a network of devices cooperates to agree on a common value~\cite{TBA, Boyd1, QuantConsensus}. Consensus is important in distributed IoT architectures because it ensures that, in computation or inference tasks, all devices agree on the result. Reference~\cite{Boyd1} studies distributed average consensus: each of the $N$ devices is assigned an initial scalar value, and their goal is to compute the average value of all of the devices. In an air quality monitoring application, a network of sensors could cooperate to find the average pollutant concentration in a city. Every device $n$ maintains a local scalar value $x_n(t)$, where $t$ is an iteration, with $x_n(0)$ equal to its initial assigned value. Again, in the air quality monitoring example, $x_n(0)$ represents the local pollutant concentration at sensor $n$. Then, every device $n$ transmits its current value or state $x_n(t)$ to all of its neighboring devices (device $n$ also receives from all its neighbors their current states) and updates its state as a weighted sum of its current state and its neighbors states
\begin{equation}\label{eqn: consensusUpdate}
	x_n(t+1) = w_{nn} x_n(t) + \sum_{j \in \Omega_n} w_{jn} x_j(t).
\end{equation}
For properly chosen weights $\left\{w_{jn}\right\}_{j, n}$, the states at each device converge toward the average of the initial data~\cite{Boyd1}.

An adversary may hijack certain devices and disrupt the consensus process. Reference~\cite{Pasqualetti1} studies distributed consensus when some devices follow an update rule that deviates from~\eqref{eqn: consensusUpdate}. That is, compromised devices do not follow~\eqref{eqn: consensusUpdate} when updating their values and instead update their values arbitrarily. To counter this adversarial behavior, the authors of~\cite{Pasqualetti1} design algorithms to detect and identify compromised devices. Recall that attack detection and identification algorithms are explicit countermeasures against adversaries. The main idea in~\cite{Pasqualetti1} is to model the consensus process as a linear dynamical system and view attacks from compromised agents as unknown inputs into the system. To detect compromised devices, an agent must determine if there is  a nonzero unknown input into the system, and to identify compromised devices, an agent must determine the locations of the nonzero attacks. From these ideas, reference~\cite{Pasqualetti1} designs algorithms for each agent $n$ to detect and identify other compromised agents using only the history of its own $x_n(t)$. These algorithms require each device to store the topology of the communication network, which becomes computationally infeasible as the number of devices in the system grows, as with IoT applications.

Reference~\cite{LeBlanc1} designs an implicit countermeasure against adversaries in consensus. In~\cite{LeBlanc1}, instead of computing the average of their initial values, the devices' goal is to simply agree on a value. That is, the devices wish to update their states such that, eventually, the uncompromised devices reach the same value. The compromised devices update their states arbitrarily and transmit falsified values to their neighbors in order to disrupt consensus. The authors of~\cite{LeBlanc1} modify the state update rule~\eqref{eqn: consensusUpdate} to deal with the compromised devices. When an agent updates its state, instead of using all of its neighbors' states, it ignores the most extreme state values. Before updating, each device $n$ sorts the states received from its neighbors $\Omega_n$ and removes the largest $F$ state values greater than its own and the smallest $F$ state values less than its own, for some predetermined number $F$. Then, each agent $n$ updates its state $x_n(t+1)$ as a weighted sum of the states from its remaining neighbors and its own current state $x_n(t)$. As long as the total number of compromised devices is less than $F$ and the communication network satisfies certain topology conditions, the modified state update rule ensures that all uncompromised agents consensus on the same state.

%\subsection{Distributed Hypothesis Testing}

\subsection{Secure Parameter Estimation: Implicit Countermeasures}
The previous subsection focused on distributed consensus, where devices converge to a common statistic from a single snapshot of their data, e.g., the average of their initial data. In distributed inference, for example, like distributed estimation, devices still converge to a common estimate of an unknown parameter, but at each communication round they make a new observation. We present three implicit countermeasures (\cite{ChenDistributed2, LeBlanc2, DiffusionLMS2}) for secure distributed inference. Reference~\cite{LeBlanc2} extends the resilient consensus algorithm in~\cite{LeBlanc1} to construct a resilient distributed estimator. In~\cite{LeBlanc2}, each device $n$ processes a stream of local measurements to recover a scalar local parameter $p_n$. In the context of air quality monitoring, the parameter $p_n$ could be the concentration of pollutants at sensor $n$. 

The authors in~\cite{LeBlanc2} consider three different types of devices to be part of the network: \begin{enumerate*} \item reliable devices, \item normal devices, and \item malicious devices. \end{enumerate*} Reliable devices directly measure their parameters of interest, and they can recover their parameters using only their local data. Normal devices are not able to directly measure their parameter. Instead, a normal device $n$ makes a relative measurement $\xi_{ln}(t) = p_l - p_n$ for every neighboring device $l$. Each device $n$ maintains an estimate $x_n(t)$, i.e., its state, of its local parameter $p_n$. At every time step, each device broadcasts its state to all its neighbors. Malicious devices may broadcast an arbitrary state or estimate. Reliable devices measure their parameter directly, so their state is the correct estimate $x_n(t) = p_n$ for all iterations. Each normal device $n$, for each of its neighbors $l \in \Omega_n$, computes a step state value
$s_{ln}(t) = x_l(t) - x_n(t) - \xi_{ln}(t).$
Device $n$ ignores the largest $F$ positive step values and the smallest $F$ negative step values. If there are fewer than $F$ positive (negative) step values, then, device $n$ ignores all positive (negative) step values. Then, each device $n$ updates its state as a weighted sum of its previous state and the remaining estimates. Reference~\cite{LeBlanc2} shows that if there are fewer than $F$ malicious devices in any device's neighborhood and if the topology of the communication network satisfies certain conditions, then, all reliable and normal devices eventually recover their parameters.

Another method for devices to deal with adversaries is to apply different gains or weights to their measurements and the information they receive from neighbors~\cite{DiffusionLMS2, ChenDistributed2}. In~\cite{DiffusionLMS2}, a network of devices make noisy measurements of an underlying parameter. The devices maintain local estimates of the parameter and update them as a weighted sum of their previous estimates (states), the states of their neighbors, and their local measurement. Malicious devices attempt to disrupt the estimation process by broadcasting false estimates to their neighbors. The authors of~\cite{DiffusionLMS2} propose an adaptive weight estimate update scheme, where an uncompromised device gives lower weight to neighbors whose estimates deviate drastically from its own. Through this adaptive weight mechanism, the uncompromised devices learn to eventually ignore malicious devices and, effectively, disconnect the adversaries from the network.

In~\cite{ChenDistributed2}, instead of hijacked devices broadcasting false estimates, the adversary attacks the network by manipulating the devices' measurements. In air quality monitoring, this corresponds to the case in which an attacker falsifies the sensor data (say., the measurement of local pollutant concentrations) on a subset of devices. The devices all observe the complete parameter and apply an adaptive gain to their measurements to mitigate the effect of the attack -- each device $n$ gives lower weight to local measurements that deviate more from its local estimate. Reference~\cite{ChenDistributed2} shows that applying lower weights to aberrant measurements limits the impact maliciously altered measurements. If less than half of the devices fall under attack, the network eventually recovers correctly the parameter of interest.

\subsection{Secure Distributed Inference: Other Work}
Further work on secure distributed inference include inference under jamming attacks~\cite{JammingAttacks} and function calculation~\cite{Sundaram1}, distributed hypothesis testing ~\cite{DistDetection, DetectionRadioactive}, and distributed optimization~\cite{Optimization1, Byrdie, Optimization2, ResilientSVM} with Byzantine agents. Reference~\cite{JammingAttacks} studies distributed estimation under jamming attacks: in jamming attacks, the adversary prevents communication between devices instead of manipulating their data streams. In~\cite{Sundaram1}, the authors design an algorithm that is resilient to Byzantines for computing a specific function of a single snapshot of data (this differs from consensus, where the goal is for the agents to converge to any common statistic).  Reference~\cite{DistDetection} studies distributed hypothesis testing with Byzantines and provides an algorithm that is resilient to a restricted class of weak Byzantine adversaries. The authors of~\cite{DetectionRadioactive} evaluate a heuristic for Byzantine-resilient distributed hypothesis testing through numerical simulations. 

In distributed optimization, each agent has a local objective function, and the agents' goal is to converge to a statistic that minimizes the sum of their objective functions, possibly subject to constraints. Reference~\cite{Optimization1} considers optimization in an all-to-all communication setup (i.e., each agent communicates with every other agent) and proposes an iterative optimization algorithm that is resilient to Byzantine agents. References~\cite{Byrdie, Optimization2} present optimization algorithms that are resilient to Byzantine agents for arbitrary network topologies. The authors of~\cite{ResilientSVM} present a Byzantine-resilient distributed optimization algorithm for training a support vector machine. 

\section{Secure Distributed Estimation Through Explicit Adversary Detection}
The algorithms provided in~\cite{LeBlanc2, DiffusionLMS2, ChenDistributed2} are implicit security countermeasures for distributed estimation: they ensure that a network of devices completes the estimation task even when adversaries attack the network, but they do not detect or identify the adversaries. We now consider an explicit countermeasure. Reference~\cite{ChenDistributed1} designs an algorithm for the subset of uncompromised devices in the network, the ones not under attack, to still simultaneously infer the value of a parameter from their stream of measurements or detect the presence of compromised devices. In a sense, this is a 0-1 strategy, the sensors surviving the attack still achieve the desired goal, or, if the attack is too strong, they are able to detect the attack and realize the presence of an intruder.

\subsection{Device Model}
Each device $n$ makes a stream of measurements, $y_n(t)$, of a parameter $\theta^*$ following
\begin{equation}\label{eqn: localMeasurement}
	\mathbf{y}_n(t) = \mathbf{H}_n \theta^* + \mathbf{w}_n(t),
\end{equation}
where $\mathbf{w}_n(t)$ is measurement noise  and the matrix $\mathbf{H}_n$ describes which parts of the parameter each device measures. For example, in air quality monitoring, $\theta^*$ represents the pollutant concentration over an entire city, and each individual component of $\theta^*$ may represent the pollutant concentration in a particular neighborhood. At each device $n$, the matrix $\mathbf{H}_n$ selects the component of $\theta^*$ corresponding to the local pollutant concentration. The parameter $\theta^*$ has bounded energy (i.e., $\left\lVert \theta^* \right \rVert \leq \eta$ for some known constant $\eta$), since, in practice, we are interested in parameters bound by physical laws. Again, in air quality monitoring, by definition, no pollutant concentration can be above $10^6$ parts per million, and in practice, the bound may be even tighter. The noise $\mathbf{w}_n(t)$ is independently and identically distributed (i.i.d.) with mean $\mathbb{E} \left[\mathbf{w}_n(t) \right] = 0$ and finite covariance $\mathbb{E} \left[ \mathbf{w}_n(t) \mathbf{w}_n(t)^\intercal \right] = \boldsymbol{\Sigma}_n$ and is independent across devices.

The network of devices, $G = (V, E)$, satisfies two natural conditions. First, the graph $G$ is connected. There is a path between any two devices, and information from each device propagates to all other devices. Second, the network of devices is \textit{globally observable}: the matrix $\sum_{n = 1}^N \mathbf{H}_n^\intercal \mathbf{H}_n$ is invertible, {\color{black}where the matrices $\mathbf{H}_n$ model the local measurement at sensor $n$ in~\eqref{eqn: localMeasurement}}. {\color{black} This global observability condition is equivalent to the rank observabilty conditions for centralized architectures.} Intuitively, global observability means that the sensors together provide meaningful information about each component of $\theta^*$. In air quality monitoring, global observability means loosely that the collective of all devices provides information about pollutant concentrations in all neighborhoods, but each individual sensor needs only to measure the local pollutant concentration of a neighborhood. %in each neighborhood, there is a sensor that is monitoring the local pollutant concentration. No individual sensor needs to measure the concentrations in \textit{all} of the neighborhoods, but, to satisfy global observability, the air quality in every neighborhood should be monitored by at least one sensor.

The goal is to recover the parameter $\theta^*$ from the measurements of the networked devices. This is to be achieved through cooperation among the devices, with each device iteratively updating their local estimate and broadcasting it to its neighbors. We assume that a subset, $\mathcal{A}$, of the devices are Byzantine, and they broadcast arbitrary estimates to their neighbors. {\color{black} The Byzantine devices are the same in each time step, i.e., the set $\mathcal{A}$ does not change over time.} The remaining uncompromised devices, $\mathcal{N}$, wish to recover $\theta^*$ even in the presence of malicious attacks of the devices in $\mathcal{A}$. 

\subsection{Distributed Estimation with Local Consistency Checks}
We now describe a resilient distributed estimation algorithm. The resilient algorithm combines a distributed detection step with distributed estimation. If the detector does not detect the presence of Byzantine actors, then, the estimation step converges with probability $1$ to the correct value $\theta^*$, even if a subset of the agents is compromised. At each time step $t = 0, 1, 2, \dots$, agent $n$ maintains a local estimate or state $\mathbf{x}_n(t)$ and a flag $\pi_n(t)$. The flag $\pi_n(t)$ is either ``Attack '' or ``No Attack'', indicating the presence (or absence) of adversaries. The algorithm iterates among three main steps: \begin{enumerate*} \item Message Passing, \item State Update, and \item Adversary Detection.\end{enumerate*} Each (uncompromised) agent $n$ initializes its state and flag as $x_n(0) = 0$ and $\pi_n(0) = \text{No Attack.}$ Compromised agents will act arbitrarily. So, here, we only specify the rules followed by the uncompromised agents. {\color{black} While compromised agents follow the policy described by the attacker, the uncompromised agents adhere to the following rules.}

\underline{Message Passing}: At time $t = 0, 1, 2, \dots$, uncompromised agent $n \in \mathcal{N}$ transmits its current state, $\mathbf{x}_n(t)$, to its neighbors.

\underline{State Update}: To average out the disturbance from the measurement noise $\mathbf{w}_n(t)$, uncompromised agent $n$ maintains a time running average of its local measurement:
\begin{equation}\label{eqn: runningMeasurement}
	\begin{split}
		\overline{\mathbf{y}}_n(t) &= {t \over t+1} \overline{\mathbf{y}}_n(t-1) + {1 \over t+1} \mathbf{y}_n(t),\\
		\overline{\mathbf{y}}_n(0) &= \mathbf{y}_n(0).
	\end{split}
\end{equation}
The uncompromised agents $n \in \mathcal{N}$ update their states following a consensus plus innovations rule (see, e.g.,~\cite{Kar2}):
\begin{equation}\label{eqn: estimateUpdate}
	\mathbf{x}_n(t+1)  = \mathbf{x}_n(t) - \underbrace{\beta \sum_{l \in \Omega_n} \left( \mathbf{x}_n(t) - \mathbf{x}_l(t) \right)}_{\text{Consensus}} + \underbrace{\alpha \mathbf{H}_n^T \left( \overline{\mathbf{y}}_n(t) - \mathbf{H}_n \mathbf{x}_n(t) \right)}_{\text{Innovations}}.
\end{equation}
The weights $\alpha$ and $\beta$ are positive weighting for the innovations and consensus terms, respectively, in~\eqref{eqn: estimateUpdate}, where the innovations term incorporates local measurements, and the consensus term propagates local measurements throughout the network and drives the agents to reach the same estimate.

\underline{Adversary Detection}: Each uncompromised agent $n \in \mathcal{N}$ checks for adversaries by comparing its own estimate with the estimates it receives from its neighbors. An agent reports the presence of adversaries if the (Euclidean) distance between its own state and the state from any of its neighbors exceeds an adaptive threshold. The uncompromised agents update their flags following
\begin{equation}\label{eqn: normalFlagUpdate}
	\pi_n(t+1) = \!\!\left\{\begin{array}{ll}\text{Attack}, & \pi_n(t) = \text{Attack}, \text{ or } \\
		& \exists l, \left \lVert x_n(t) - x_l(t) \right\rVert > \gamma_t \\
		\text{No Attack}, & \text{Otherwise} \end{array}\right.,
\end{equation}
where $\gamma_t$ is a time-varying adaptive threshold. The threshold $\gamma_t$ follows the recursion
\begin{equation}\label{eqn: gammaDef}
\begin{split}
	\gamma_{t+1} &= \underbrace{\left(1 - r_1 \right)\gamma_t}_{\text{Error Buffer}} + \underbrace{\alpha\frac{2K}{(t+1)^{\tau}}}_{\text{Noise Buffer}}, \\
	\gamma_0 &= 2\eta \sqrt{N},
\end{split}
\end{equation}
and depends on the parameters $K > 0$, $0 < \tau < \frac{1}{2}$, and $0 < r_1 \leq 1$. Recall that $\eta$ bounds the energy of the parameter $\theta^*$, and $N$ is the total number of devices.

The threshold $\gamma_t$ describes how far apart the states of two neighboring devices should be if there is no adversary. It consists of two components: the error buffer component, $(1-r_1) \gamma_t$, and the noise buffer component, $\alpha {2K \over (t+1)^\tau}$. As the agents follow the state update, their states move closer to those of their neighbors. The error buffer describes the rate at which neighboring devices' states move closer together in the absence of adversaries and noise. The noise buffer compensates for the effect of measurement noise and depends on the parameters $K$ and $\tau$. The parameter $K$ describes the base size of the noise buffer at each iteration, and the parameter $\tau$ describes how the noise buffer decays over time. If the threshold $\gamma_t$ is too small (e.g., if base size $K$ of the noise buffer is too small), then, the algorithm incurs a high probability of false alarm, since measurement noise may cause neighboring agents' states to exceed threshold.

Adversarial devices update their own estimates in an arbitrary manner and have no need for a flag. To avoid detection, adversarial agents, which, in the extreme case we assume, know all algorithm parameters, must ensure that, for all times $t$, the distance between the state they transmit and each of their neighbors' states is less than the threshold $\gamma_t$. There is a tradeoff between the magnitude of the threshold and the performance of the algorithm. For large threshold values, the algorithm has few false alarms, but adversarial devices may transmit estimates that deviate more significantly while evading detection. Small thresholds detect adversaries more effectively but suffer from more false alarms.

The beauty of the approach in~\cite{ChenDistributed1} is that by careful design of the algorithm parameters $\alpha$, $\beta$, and $\gamma_t$, one can guarantee (see the Estimator Performance section below) that either an attack is detected or the estimator is accurate. The parameters $K$, the base size of the noise buffer, and $\tau$, the decay rate of the noise buffer, may take any values that satisfy $K >0$ and $0 < \tau < {1 \over 2}$. For the agents to effectively recover $\theta^*$ and detect adversaries with low false alarm probability, we must choose $\alpha$, the innovation weight, $\beta$, the consensus weight, and $r_1$, the decay rate of the error buffer, to satisfy certain eigenvalue conditions related to the dynamics of the estimate update rule~\eqref{eqn: estimateUpdate}. The algorithm in~\cite{ChenDistributed1} is fully distributed and requires only local data at each agent. This differs from attack detectors for architectures with central processors (e.g.,~\cite{Liu}), which require the central processor to have access to all data streams. Additionally, the algorithm in~\cite{ChenDistributed1} does \textit{not} require each agent to store the topology of the network locally, unlike the attack detector for consensus algorithms presented in~\cite{Pasqualetti1}. 

\subsection{Estimator Performance}
In the absence of Byzantine agents, the algorithm from~\cite{ChenDistributed1} ensures that all agents produce strongly consistent estimates (i.e., they eventually recover the parameter $\theta^*$ almost surely) and has few false alarms. The false alarm rate can be made arbitrarily small by choosing a larger noise buffer base size $K$. When there are Byzantine agents, the performance of the algorithm depends on the distributed observability of the remaining, uncompromised agents, $\mathcal{N}$. Consider the network of uncompromised agents only, and suppose that this network is connected and globally observable. In the presence of adversarial agents, one of two events must occur: either \begin{enumerate*} \item at some time $t$ an uncompromised agent $n$ changes its flag value to $\pi_n(t) = \text{Attack,}$ or  \item no uncompromised agent ever changes its flag value. \end{enumerate*} If the first event occurs, the algorithm successfully detects the Byzantine agents.
%\begin{theorem}\label{thm: regularConvergence}
%	If there are no adversarial agents, then, under the algorithm in~\cite{ChenDistributed1}, for all agents $n \in \left\{1, 2, \dots, N\right\}$, we have
%\begin{equation}\label{eqn: regularConvergence}
%	\mathbb{P} \left( \lim_{t \rightarrow \infty} \left(t+1\right)^{\tau_0} \left\lVert \mathbf{x}_n(t) - \theta^* \right\rVert_2 = 0 \right) = 1,
%\end{equation}
%for every $0 \leq \tau_0 < {1 \over 2}$.
%Moreover, the false alarm probability, $P_{FA}$, satisfies
%\begin{equation}\label{eqn: falseAlarm}
%	P_{FA} \leq \frac{\Psi \zeta(\tau)}{K^2},
%\end{equation}
%where $\Psi = \sum_{n = 1}^N \trace\left(\boldsymbol{\Sigma}_n\right)$ and $\zeta(\tau) = \sum_{j = 1}^\infty \frac{1}{j^{2\left(1-\tau\right)}}$.
%\end{theorem}

%\noindent The agents' local estimates converge almost surely to the true parameter value $\theta^*$ at the rate of ${1 \over t^{\tau_0}}$ for any $0 < \tau_0 < \frac{1}{2}$. 

If the second event occurs, the adversarial agents evade detection. To evade detection, each adversarial agent $n$ may only transmit states that deviate from their neighbors' states by less than the threshold, $\gamma_t$. The adaptive threshold decays over time, which means that, to evade detection, the adversarial agents' attack must become weaker over time. Under the conditions of connectivity and global observability, the network of uncompromised agents still produce consistent estimates when the adversarial agents evade detection. In this sense, the algorithm from~\cite{ChenDistributed1} outperforms standard anomaly detectors for distributed architectures (e.g.,~\cite{Pasqualetti1}): the distributed attack detector guarantees that, if there is a missed detection, then the agents still produce consistent estimates. Standard anomaly detectors provide no such guarantee. The key conditions for resilience under the algorithm from~\cite{ChenDistributed1} are that the network of uncompromised devices is connected and globally observable. If these conditions are not satisfied, then, it is possible for the adversaries to disrupt the estimation process (i.e., cause the devices to produce inconsistent estimates) while evading detection.

%\begin{theorem}\label{thm: resilientOperation}
%	Let the network of normally behaving agents $\mathcal{N}$ be connected and globally observable. If, for all $n \in \mathcal{N}$ and for all $t = 0, 1, \dots$, we have $\pi_n(t) = \text{No Attack}$, for all $n \in \mathcal{N}$, we have
%\begin{equation}\label{eqn: resilientConvergence}
%	\mathbb{P} \left( \lim_{t \rightarrow \infty} \left(t+1\right)^{\tau_0} \left\lVert \mathbf{x}_n(t) - \theta^* \right\rVert_2 = 0 \right) = 1,
%\end{equation}
%for every $0 \leq \tau_0 < {\tau}$.
%\end{theorem}
%\noindent When adversaries evade detection, the uncompromised agents' local estimates converge almost surely to the true parameter value $\theta^*$ at the rate of ${1 \over t^{\tau_0}}$ for any $0 < \tau_0 < \tau$. In the presence of adversaries, the estimate convergence rate is limited by $\tau$, the decay parameter of the noise buffer.

\subsection{Numerical Example}
\begin{figure}[h!]
	\centering
	\includegraphics[keepaspectratio = true, scale = .6]{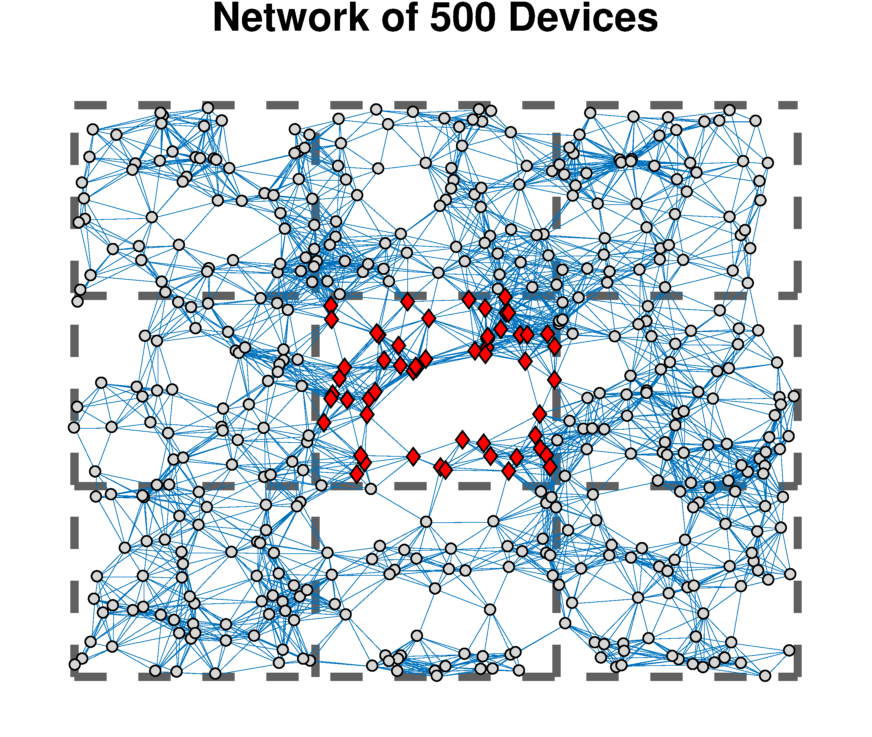}
	\caption{Network of $N = 500$ sensors. Each sensor measures pollutant concentrations in its local sector only. An adversary hijacks a subset of sensors in the center sector, denoted by red diamonds.}
	\label{fig: network}
\end{figure}

As an illustration, consider air quality monitoring in smart cities. For example, the city of Chicago plans to deploy $500$ sensors by the end of 2018 to monitor environmental conditions~\cite{AoT}. Figure~\ref{fig: network} shows a network of $N = 500$ sensors deployed in $9$ sectors of a city. Sensors are placed uniformly at random over a 2 kilometer by 2 kilometer grid. Two sensors share a communication link if they are located within 200 meters of each other. Each sensor measures the pollutant concentration in its own sector only, and their goal is to recover the pollutant concentrations over all $9$ sectors. 

Each component of $\theta^*$, representing local pollutant concentrations, is drawn independently and uniformly between $0$ $\mu \text{g}/\text{m}^3$ and $160$ $\mu \text{g}/\text{m}^3$.\footnote{According to the United States Environmental Protection Agency, the maximum safe level of Particulate Matter 10 ($\text{PM}_{10}$) is $150$ $\mu \text{g}/\text{m}^3$~\cite{EPA}.} Each device's sensor is corrupted by additive Gaussian white noise with mean $0$ and variance $\Sigma_n = 10$. The local signal-to-noise ratio (SNR) is $13 \text{ dB}$. We demonstrate the performance of the algorithm in three different scenarios: \begin{enumerate}\item \textbf{No Adversaries}: All devices remain uncompromised. \item \textbf{Strong Adversaries}: An adversary compromises all devices in the center sector. The remaining devices are no longer globally observable. \item \textbf{Weak Adversaries}: An adversary compromises half of the devices in the center sector. The remaining devices are connected and globally observable. \end{enumerate}. 

\begin{figure}[h!]
\centering
\begin{subfigure}[h]{0.45\columnwidth}
	\includegraphics[width = \columnwidth]{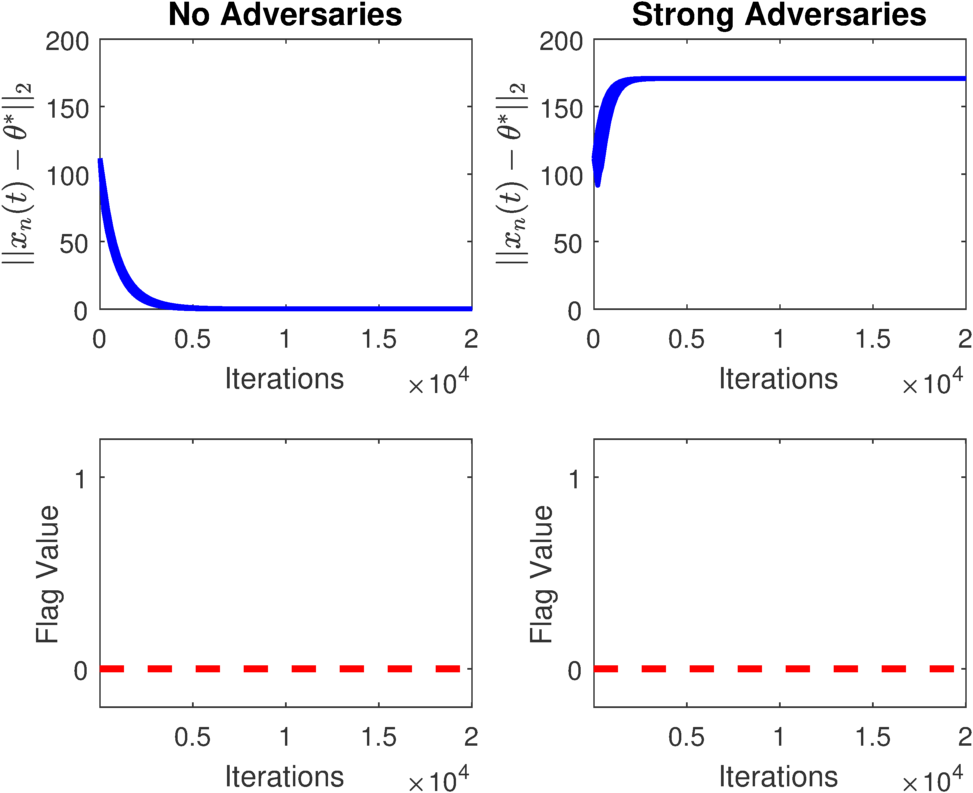}
	\caption{}\label{fig: perf1}
\end{subfigure}
\begin{subfigure}[h]{0.45\columnwidth}
	\includegraphics[width = \columnwidth]{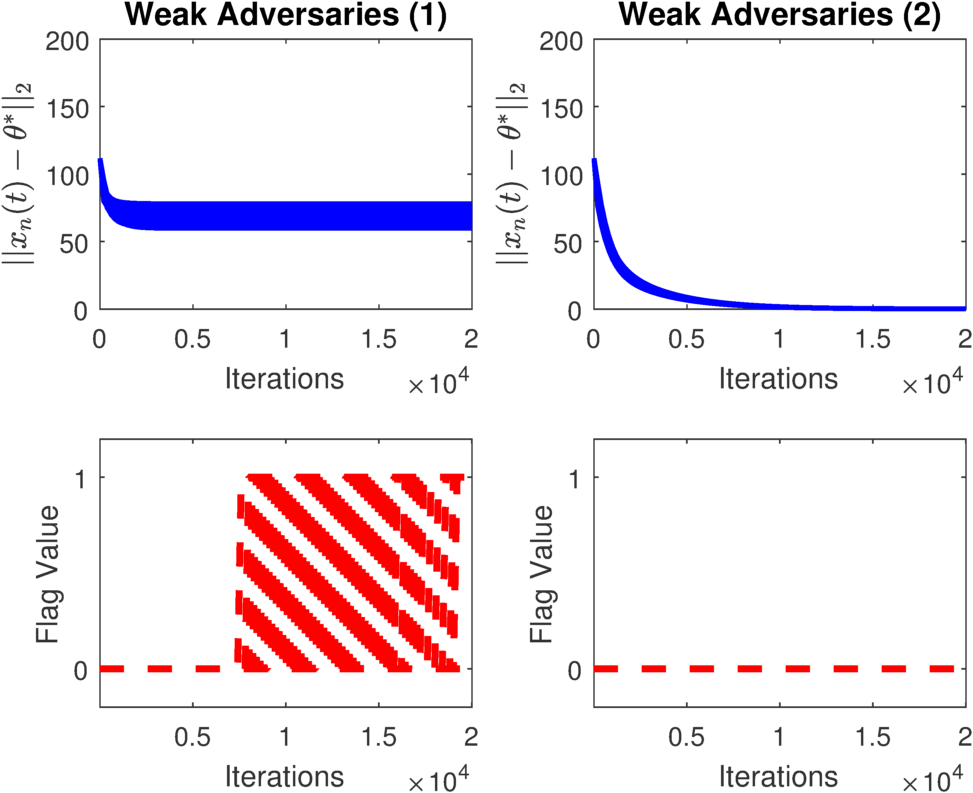}
	\caption{}\label{fig: perf2}
\end{subfigure}
\caption{Evolution of estimation errors and flag values over iterations of~\eqref{eqn: estimateUpdate} and~\eqref{eqn: normalFlagUpdate} for uncompromised devices with (a) no adversaries, strong adversaries, and (b) weak adversaries. When an agent detects an adversary, it changes its flag value from $0$ to $1$. }\label{fig: Performance}
\end{figure}

Figure~\ref{fig: Performance} depicts the performance of the algorithm and shows the evolution of the agents' estimation errors and flag values over $20,000$ iterations. When there are no compromised devices, the estimates of all devices converge to $\theta^*$, and no device reports the presence of adversaries. 
When all of the devices in the center sector are compromised, the remaining devices are unable to recover $\theta^*$ and do not detect the adversaries. This is because the network of remaining uncompromised devices is not globally observable. It has no information about the pollutant concentrations in the center sector. When only half of the devices in the sector are compromised, the network of uncompromised devices is globally observable. In this case, adversaries that disrupt the estimation process are eventually detected. If adversaries attempt to evade detection, then, the remaining devices eventually recover the global pollutant concentration, $\theta^*$, although, in this case, the devices' estimates converge more slowly compared to the case where there are no adversaries.

\section{Conclusion}
In this paper, we presented an overview of methods for resilient decentralized and distributed inference in the Internet of Things. We have separately considered explicit countermeasures, such as adversary detection and identification algorithms, and implicit countermeasures, inference algorithms that are inherently resilient to data manipulation. A general requirement for achieving resilience is that the uncompromised, cooperative devices have enough influence to overcome the disruptive effects of adversarial devices. In simple settings, e.g., where all devices observe the same phenomena, this means that a majority of devices should be uncompromised. In settings where devices observe different phenomena, for example, different components of an unknown parameter, resilience depends on the observability of the uncompromised devices.

There are several open challenges for secure distributed inference in the IoT. First, for fully distributed IoT systems, we have focused primarily on static inference tasks, e.g., estimating a parameter that does not change (or changes slowly) over time. It is also necessary to design countermeasures for dynamic distributed inference tasks where the target parameter changes quickly in time or where agents move and the network changes over time, e.g., a network of automobiles estimating traffic conditions. In cases where agents are mobile, an adversarial agent may move into different agents' neighborhoods over time, making the problem of detecting and identifying adversaries more challenging.

Second, we have focused on inference tasks where all of the devices aim to recover the same parameter or decision. Another area of future work is designing resilient algorithms for inference tasks where devices have different goals. For example, in air quality monitoring, a device may be interested in recovering the pollutant concentration in its sector and nearby sectors only instead of recovering the pollutant concentrations over an entire city. 

Finally, we have focused on countermeasures that ensure systems \textit{complete} their inference tasks. Depending on the adversary this may not be possible, e.g., if, in decentralized hypothesis testing, the majority of devices are compromised. A goal of future work is to design countermeasures that ensure graceful performance degradation when complete resilience is not achievable. 

\section*{Acknowledgments}
Figures and diagrams are courtesy of Jasper Tom.

\bibliography{IEEEabrv,References}
\vspace{-.25 in}
\begin{IEEEbiographynophoto}{Yuan Chen}(S'14) received the B.S.E. degree in electrical engineering from Princeton University, Princeton, NJ, in 2013. Since 2013, he has been pursuing the Ph.D. degree in the electrical and computer engineering at Carnegie Mellon University, Pittsburgh, PA. His current research activities are focused on distributed inference, cyber-physical systems (CPS), and security for the Internet of Things (IoT).
\end{IEEEbiographynophoto}
\vspace{-.25 in}
\begin{IEEEbiographynophoto}{Soummya Kar}
(S'05--M'10)  received a B.Tech. in electronics and electrical communication 
engineering from the Indian Institute of Technology, Kharagpur, India, in May 2005 and 
a Ph.D. in electrical and computer engineering from Carnegie Mellon University, 
Pittsburgh, PA, in 2010. From June 2010 to May 2011, he was with the Electrical Engineering Department, Princeton University, Princeton, NJ, USA, as a Postdoctoral Research Associate. He is currently an Associate Professor of Electrical and Computer Engineering at Carnegie Mellon University, Pittsburgh, PA, USA. His research interests include decision-making in large-scale networked  systems, stochastic systems, multi-agent systems and data science, with applications to cyber-physical systems and smart energy systems.
\end{IEEEbiographynophoto}
\vspace{-.25 in}
\begin{IEEEbiographynophoto}{Jos\'e M.~F.~Moura}(S'71--M'75--SM'90--F'94) received the engenheiro electrot\'{e}cnico degree from Instituto Superior T\'ecnico (IST), Lisbon, Portugal, and the M.Sc., E.E., and D.Sc.~degrees in EECS from the Massachusetts Institute of Technology (MIT), Cambridge, MA.

He is the Philip L.~and Marsha Dowd University Professor at Carnegie Mellon University (CMU). He was on the faculty at IST and has held visiting faculty appointments at MIT and New York University (NYU). He founded and directs a large education and research program between CMU and Portugal, www.icti.cmu.edu.

His research interests are on data science, graph signal processing, and statistical and algebraic signal and image processing. He has published over 550 papers and holds fourteen patents issued by the US Patent Office. The technology of two of his patents (co-inventor A. Kav\v{c}i\'c) are in about three billion disk drives read channel chips of 60~\% of all computers sold in the last 13 years worldwide and was, in 2016, the subject of the largest university verdict/settlement in the information technologies area.

Dr. Moura is the 2018 IEEE President Elect. He has received several awards, including  the Technical Achievement Award and the Society Award from the IEEE Signal Processing Society. In 2016, he received the CMU College of Engineering Distinguished Professor of Engineering Award. He is a Fellow of the IEEE, a Fellow of the American Association for the Advancement of Science (AAAS), a corresponding member of the Academy of Sciences of Portugal, Fellow of the US National Academy of Inventors, and a member of the US National Academy of Engineering.
\end{IEEEbiographynophoto} 

\end{document}